\documentclass[twocolumn,preprint,aps,floats,nofootinbib,showpacs]{revtex4-1}
\usepackage{float,subfig,graphicx,epsfig,epsf}
\usepackage{multirow}
\usepackage{url}
\usepackage{amsmath,units}
\usepackage[usenames,dvipsnames,svgnames,table]{xcolor}
\usepackage[colorlinks=true,linkcolor=red,urlcolor=blue,citecolor=blue]{hyperref}

\begin{document}

	\thispagestyle{empty}

	\title{
		A comparative study of Compact Objects using 3 models: 	
		Walecka Model, PAL Model, and M.I.T. Bag Model
		}

	\author{
		F\'{a}bio Kopp$^{1}$\footnote{email: fabio.kopp@ufrgs.br}, 
		Alex Quadros$^{2}$, 
		Guilherme Volkmer$^{1}$, 
		Mois\'{e}s Razeira$^{3}$, 
		Magno Machado$^{1}$, 
		Dimiter Hadjimichef$^{1}$, 
		C\'{e}sar Augusto Zen Vasconcellos$^{1}$\footnote{email: cesarzen@if.ufrgs.br}
		}

	\affiliation{
		 {\small $^{1}$Instituto de F\'{i}sica, Universidade Federal do Rio Grande 
		 do Sul, Caixa Postal 15051, CEP 91501-970, Porto Alegre, RS, Brazil}  \\
		 {\small $^{2}$ High and Medium Energy Group, Instituto de F\'{i}sica e 
		 Matem\'{a}tica, Universidade Federal de Pelotas, Caixa Postal 354, 
		 CEP 96010-900, Pelotas, RS, Brazil }\\
		 {\small $^{3}$Instituto de F\'{i}sica, Universidade Federal do Pampa, 
		 CEP: 96570-000,Ca\c{c}apava do Sul, RS, Brazil}
		 }

	\begin{abstract}
		Compact objects are the name used to classify the following objects: white dwarf, neutron star and black holes. In addition to them, some authors have suggested that in the core of a neutron star we may have the quarks u,d,s and c. In this work we investigate the structure 
		of neutron star and strange star, described by the Tolman-Openheimer-Volkof equations. For the neutron star, we used the relativistic EoS from Walecka's model and  the non-relativistic EoS PAL model. For the strange star, we used the M.I.T. bag model.
		We obtain results for the mass-radius relation and then we compared 
		ours results with the actual pulsar data recently observed PSR J1614-2230 
		with a mass $\unit[1.97\pm 0.04]{M_{\odot}}$.
	\end{abstract}

 
  \maketitle
  
	\section{Introduction}
	
	Motivated by the recent measurement of the Shapiro delay in the radio pulsar 
	PSR J1614--2230 yielded a mass of $\unit[1.97\pm 0.04]{M_{\odot}}$ 
	\cite{pulsar}, we have selected two models based on the Equation of State 
	(EoS) of dense matter namely Walecka model \cite{walecka}, M. Prakash, T.L. Ainsworth, and 
	J.M. Lattimer model \cite{pal}. For the study of strange star we have selected the M.I.T Bag model \cite{mit}. We used these EoS to 
	solve the Tolman-Oppenheimer-Volkov equations (TOV) \cite{tov} for compact 
	stars. Then we compare the mass obtanied with the PSR J1614--2230 mass.
	
 \section{Formalism}


	Assuming spin-1/2 nucleons coupled to isoscalar scalar $\sigma$-meson and the 
	isoscalar vector $\omega$-meson, the EoS of Walecka model \cite{walecka} is
	obtained from the ground state expectation value of the time and space 
	components of the diagonal energy-momentum tensor and reads
  \begin{equation}
  \begin{aligned}
		\epsilon =\frac{g_{v}^{2}}{2m_{v}^{2}}\rho_{B}^{2} + 
		\frac{m_{s}^{2}}{2g_{s}^{2}}(M - M^{*})^{2} + \\ 	
		\frac{\gamma}{ (2\pi)^{3} } \int_{0}^{k_{F}}d^{3}k\sqrt{k^2+M^{*2}}
    \end{aligned}
	\end{equation}

	\begin{equation}
	\begin{aligned}  
	p =\frac{g_{v}^{2}}{2m_{v}^{2}}\rho_{B}^{2} - 
	\frac{m_{s}^{2}}{2g_{s}^{2}}(M - M^{*})^{2} + \\
	\frac{1}{3}\frac{\gamma}{ (2\pi)^{3} } \int_{0}^{k_{F}}d^{3}k
	\frac{k^2}{\sqrt{k^2+M^{*2}}}
    \end{aligned}
     \end{equation}
	where $\rho_{B} =\frac{\gamma}{6\pi^{2}}k_F^3$ is the nuclear density, 
	$\gamma=4$ is the degeneracy factor of the nuclear matter, and $k_{F}$ is the 
	Fermi momemtum. The meson-nucleon coupling constants are $g_s$ and $g_v$; 
	$m_v$ and $m_s$ are the	masses of the mesons. The quantity,
	$M^{*}=M-\left(\frac{g_{s}}{m_{s}}\right)^{2}\frac{\gamma}{(2\pi)^{3}}
	\int_{0}^{k_{F}}d^{3}k\frac{M^{*}}{\sqrt{k^2+M^{*2}}}$, is the effective mass and $M$ is the nucleon mass.


	
 The PAL model \cite{pal} follows the same strategy applied in Walecka model, 
 since its values for the saturation point are fitted to reproduce the experimental 
 values. In this context, the EoS is given by
\begin{equation}
  \begin{aligned}  
  	\epsilon =\rho_{0}u
  	\Bigg\{m_n+E_F^{(0)}u^{2/3}+\frac{Au}{2}+
  	\frac{Bu^{\sigma}}{1+B'u^{\sigma-1}}+ \\ 3\sum_{i=1}^{2}C_{i}
  	\left(\frac{\Lambda_{i}}{p_{F}^{(0)}}\right)^3
  	\left[\frac{p_{F}}{\Lambda_{i}}-
  	\tan^{-1}\left(\frac{p_{F}}{\Lambda_{i}}\right)\right]
  	\Bigg\},\\
  \end{aligned}
  \end{equation}

\begin{equation}
\begin{aligned}   
  	p =\rho_{b}\frac{d\epsilon}{d\rho_{b}}-\epsilon.
  \end{aligned}
  \end{equation} 
  The saturation density is $\rho_{0}=\unit[0.16]{fm^{-3}}$ and 
  $u=\frac{\rho_b}{\rho_0}$ with $\rho_{b}=\frac{k_{F}^{3}}{3\pi^2}$. The 
  symmetry energy is defined as $S=\left(2^{2/3}-1\right)\frac{3}{5}E_{F}^{(0)}
  \left(u^{2/3}-F(u)\right)+ S_{0}F(u)$. The Fermi energy  at saturation point is $E_F^{0}=-16$ MeV.
  


  In the M.I.T bag model \cite{mit}, the quarks up ($u$), down ($d$) and strange 
  ($s$) are inside a bag which has a constant positive potential per unit of 
  volume, the so-called Bag constant ($B$). The EoS for this model is

 \begin{equation}
 \begin{aligned}
 		&\epsilon=B  \ + \\ \sum_{u,d,s} \frac{3}{4\pi^2}
 		&\left[\mu_f k_f
 				\left(\mu_f^2-\frac{1}{2}m_{f}^2\right)  - 
 				\frac{1}{2}m_f^4\ln\left(\frac{\mu_f+k_f}{m_f}\right)\right]
  \end{aligned}
  \end{equation}
 \begin{equation}
 \begin{aligned}  
  &p =-B \ +  \\  \sum_{u,d,s} \frac{1}{4\pi^2}
  &\left[
  \mu_f k_f
  \left(\mu_f^2-\frac{5}{2}m_{f}^2\right)+
  \frac{3}{2}m_f^4\ln\left(\frac{\mu_f+k_f}{m_f}\right)
  \right]
  \end{aligned}
  \end{equation}
 where the chemical potential is defined as $\mu_f=\sqrt{m_f^2 + k_f^2}$. For 
 this model we will assume $u$, $d$, and $s$ quarks; and the bag constant is set 
 to be $B^{1/4}=\unit[145]{MeV}$.



	Before we finalize the methodology of study we present the structure equations
	used to calculate the masses and radii of these compact objects. The TOV equations for 
	an isotropic, general relativistic, static, ideal fluid sphere in hydrostatic 
	equilibrium reads
 	 \begin{equation}
 	 \begin{aligned} 
		&\frac{dP}{dr} = -\frac{G\epsilon(r)M(r)} {c^{2}r^{2}}
		\left[ 1 + \frac{P(r)}{\epsilon(r)} \right]\\
		&\left[1 + \frac{4 \pi r^{3} P(r)} {M(r) c^{2}} \right]
		\left[1 - \frac{2 G M(r)} {c^{2} r} \right]^{-1},
		 \end{aligned}
		 \end{equation}
		 
		 \begin{equation}
		 \begin{aligned}
		\frac{dM}{dr} &= \frac{4 \pi r^{2} \epsilon(r)} {c^{2}},
	 \end{aligned}
	 \end{equation}
	where $c$ is the is the speed of light in vacuum. The set of differential 
	equations is subject to the initial conditions $P(r=0)=P_{0}$, $P(r=R)=0$; 
	$M(r=0)=0$ and $M(r=R)=M$.	

 \section{Results and conclusions}

\begin{figure}[!htp]
		\captionsetup{type=figure}
			\begin{center}
				\subfloat[{\small Walecka Model, PAL Model and M.I.T. Bag Model EoS,
				respectively, compared with causal limit (dashed line).}]
				{\label{fig:1}
				\includegraphics[scale=0.6]{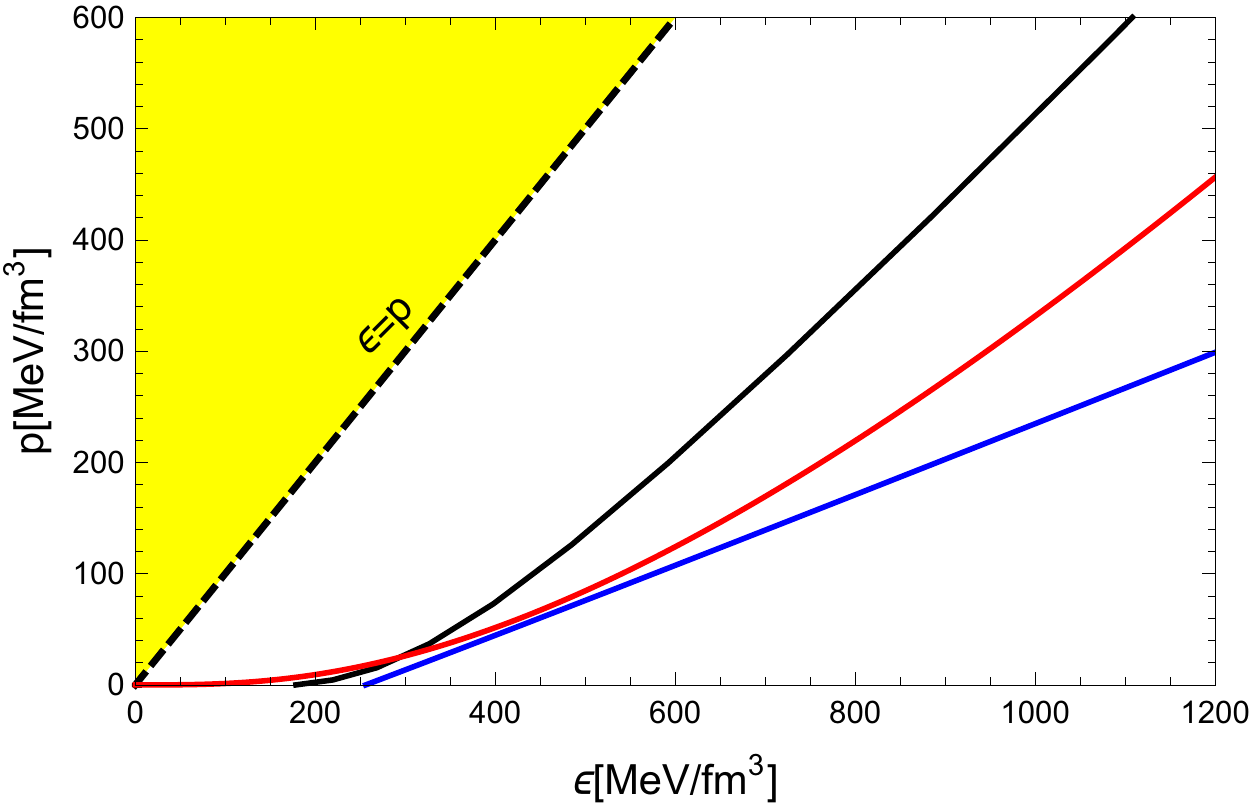} }
				\;
				\subfloat[{\small Numerical solution of TOV equations. The horizontal
				lines represents the mass 	interval observed for PSR J1614-2230.}]
				{\label{fig:2}
				 \includegraphics[scale=0.6]{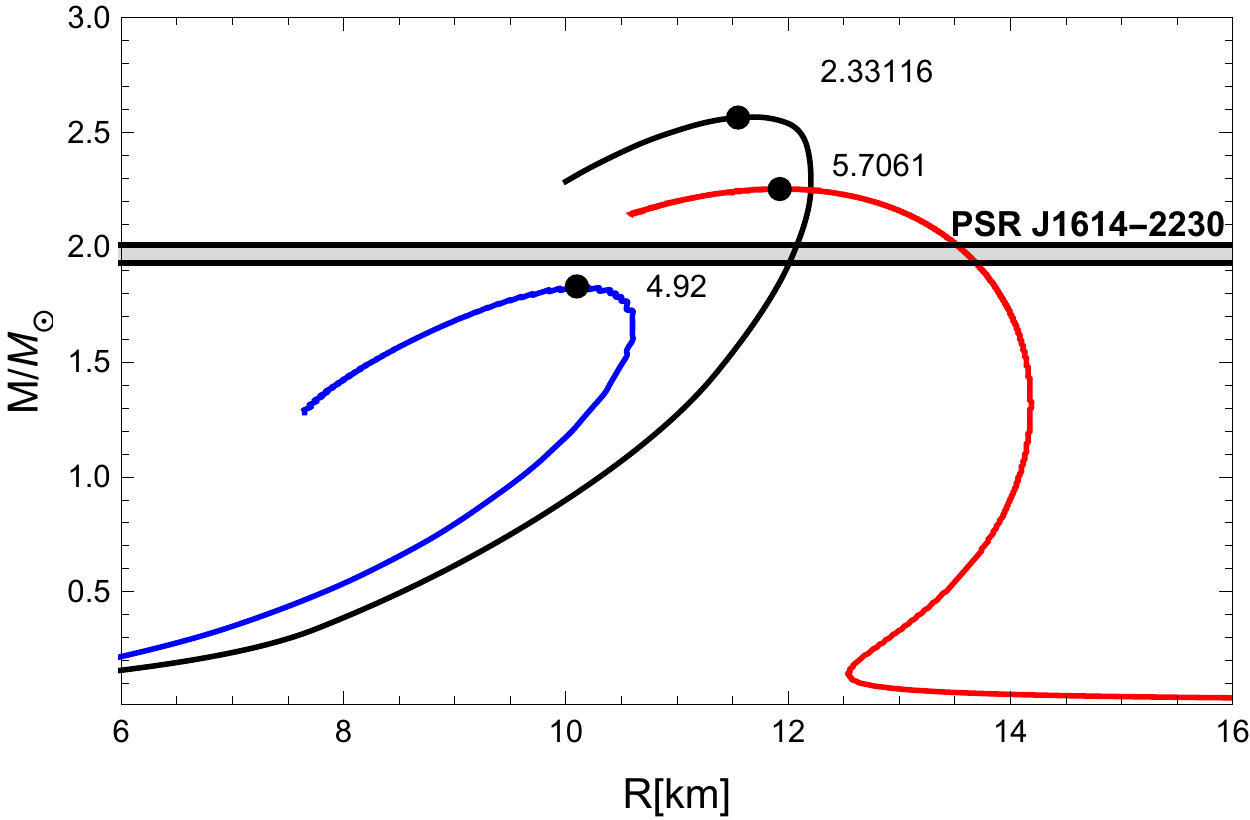} }
		\captionof{figure}{\small In both FIG. \ref{fig:1} and FIG. \ref{fig:2} 
		we set the black color to Walecka Model, red color to PAL Model 
		and blue color to M.I.T. Bag Model. The black dots in the FIG. \ref{fig:2}
		represents the maximum value reached by the model and its density at that 
		point.}
     \label{plot:1}
			\end{center}
	\end{figure}

	The EoS parameters of the three models discussed in the present work 
	FIG. \ref{plot:1} can be found in the following references \cite{walecka,pal,mit}. 
	In the FIG. \ref{fig:1} we show a comparative graph between the three EoS and the 
	causal limit ($\epsilon=p$). Then we solved the TOV equations numerically 
	using the 4th Order Runge-Kutta method and present the results of each TOV
	integration with differents EoS in the FIG. (\ref{fig:2}).
	
	In sumary the stellar macroscopic properties obtained in the present work are
	$M/M_\odot=2.65$, $R=\unit[11.55]{km}$, and $\rho/\rho_{0}=2.33$ for Walecka
	model; $M/M_\odot=2.25$, $R=\unit[11.92]{km}$, and $\rho/\rho_{0}=5.70$ for 
	PAL model, and finaly, $M/M_\odot=1.82$, $R=\unit[10.11]{km}$, and 
	$\rho/\rho_{0}=4.92$ for M.I.T bag model.
%
%
	
  The nuclear models presented in this work yields maximum masses higher 
  than the actual observation of the pulsar PS1614-2230, while the 
  result of M.I.T bag model (quark matter) does not reach the 
  $\unit[1.97]{M_{\odot}}$ mass barrier. However, the realistic approach 
  must consider the EoS that agrees with all the symmetric nuclear 
  properties, a modified version of TOV to include rotation and strong 
  magnetic field. This will be the next step of the present work.
  	
	
\section{Acknowledgments}
	We would like to thank CAPES (Coordena\c{c}\~{a}o de Aperfei\c{c}oamento de 
	Pessoal de N\'{i}vel Superior) and CNPq for the financial support.


\end{document}